\documentclass[aps,prb,twocolumn,floatfix,showpacs,showkeys,amssymb]{revtex4}

\usepackage{graphicx}

\begin{document}

\date{\today} 

\def\ba{\begin{array}}
\def\ea{\end{array}}
\def\be{\begin{equation}\begin{array}{l}}
\def\ee{\end{array}\end{equation}}
\def\bea{\begin{equation}\begin{array}{l}}
\def\eea{\end{array}\end{equation}}
\def\f#1#2{\frac{\displaystyle #1}{\displaystyle #2}}
\def\om{\omega}
\def\omm{\omega^a_b}
\def\we{\wedge}
\def\de{\delta}
\def\De{\Delta}
\def\va{\varepsilon}
\def\omb{\bar{\omega}}
\def\la{\lambda}
\def\vv{\f{V}{\la^d}}
\def\si{\sigma}
\def\t{T_+}
\def\v{v_{cl}}
\def\m{m_{cl}}
\def\n{N_{cl}}
\def\bi{\bibitem}
\def\c{\cite}
\def\sa{\sigma_{\alpha}}
\def\ua{\uparrow}
\def\da{\downarrow}
\def\mua{\mu_{\alpha}}
\def\ga{\gamma_{\alpha}}
\def\g{\gamma}
\def\ora{\overrightarrow}
\def\pa{\partial}
\def\ov{\ora{v}}
\def\al{\alpha}
\def\bt{\beta}
\def\R{R_{eff}}
\def\th{\theta}

\def\muu{\f{\mu}{ed}}
\def\E{\f{edE(\tau)}{\om}}
\def\t{\tau}

\title{Extended states in disordered systems:
role of off-diagonal correlations}

\author{Wei Zhang and Sergio E. Ulloa}

\affiliation{Department of Physics and Astronomy, and Nanoscale
and Quantum Phenomena Institute, Ohio University, Athens, Ohio
45701-2979}

\begin{abstract}

We study one-dimensional systems with random diagonal disorder but
off-diagonal short-range correlations imposed by structural
constraints. We find that these correlations generate effective
conduction channels for finite systems. At a certain golden
correlation condition for the hopping amplitudes, we find an
extended state for an infinite system. Our model has important
implications to charge transport in DNA molecules, and a possible
set of experiments in semiconductor superlattices is proposed to
verify our most interesting theoretical predictions.

\end{abstract}

 \pacs{72.80.Le, 71.55.Jv, 85.65.+h, 72.15.Rn}
 \keywords{disorder, correlation, molecular electronics}

\maketitle

Electronic states in disordered system have been an active
research topic for many years. Ever since the pioneering work of
Anderson, \c{scale} it has been generally believed that disorder
in low dimensional systems leads to unequivocal localization of
electrons. However, the situation changes if additional structure
or correlations are imposed on the statistical properties of the
randomness. It was found, for example, that a few special extended
states in a 1D ``random dimer" model exist due to symmetries of
the resonant scattering in the structure. \c{ram} Furthermore, the
existence of a mobility edge separating extended and localized
states was confirmed for 1D random systems with weak long-range
{\em correlated} disorder. \c{long}

Most studies have concentrated on diagonal disorder, where the
local energies in a tight-binding description are assigned
randomly, although some studies have explored the role of
off-diagonal disorder, where the intersite hopping constants are
chosen from a random distribution.  The role of correlated
diagonal and off-diagonal disorder has received attention only
recently, both theoretically, \c{von,nancy} and experimentally.
\c{hei} Moreover, in many systems, local correlations appear
naturally due to the built-in chemical structure. In this article,
we investigate the effects of structural constraints on the
correlated diagonal and off-diagonal disorder, and their impact on
charge transport. We find that the local correlations generate
extended states, which therefore enhance electronic transport even
in the macroscopic limit.

Our studies have been motivated in part by questions on the nature
of charge transport in DNA, a subject which has arisen much
interest recently, due to its fundamental roles in biological
processes and in possible novel device designs. \c{dna} A DNA
molecular system can be viewed as a 1D chain composed of base
pairs AT and CG in a typically random order. The on-site energies
for pairs of bases AT and CG are different, corresponding to the
different ionization potentials. \c{conwell} The role of onsite
energy correlations in different DNA sequences was discussed
recently in the literature, \c{carpena} where a constant hopping
amplitude was considered. However, the hopping amplitude (via
$\pi$-orbital overlap) in DNA molecules depends on whether the
electron (or hole) hops between AT/AT, AT/CG, or CG/CG base pairs.
\c{endres} The short-range correlation of the hopping amplitudes
due to the built in chemical structure is shown to affect the
transport properties and effectively open conduction channels for
electrons in DNA molecules -- even in those with fully random
sequences, such as $\la$-phage DNA. The transport properties are
shown to be actually determined by a subtle competition between
the disorder in base pair arrangement (onsite disorder) and
hopping (``off-diagonal") correlations.

The minimal model to study random systems with diagonal and
off-diagonal disorder is an  effective 1D tight-binding model
described by the Hamiltonian \be H=\sum_j [\va_j c_j^+c_j +
t_{j,j+1}(c_j^+ c_{j+1}+c^+_{j+1}c_j)], \ee where the onsite
energies are chosen from the bivalued distribution $\va_j=\va_A$
and $\va_B$. Correspondingly, the hopping constants are given by
$t_{j,j+1}=t_{AA}$ (or $t_{BB}$), if $\va_j=\va_{j+1}=\va_A$ (or
$\va_{BB}$); while $t_{j,j+1}=t_{AB}$, otherwise. This model is
perhaps the simplest generalization of the Anderson model, which
is the limit for $t_{AA}=t_{BB} =t_{AB}$. Notice that in DNA, the
$A$ and $B$ labels refer to the two kinds of base pairs, AT and
CG, while the model could be easily adapted to describe electronic
states in other complex molecules (polymers) and/or semiconductor
superlattices, as we will discuss below. \footnote{$\pi$-orbital
chains for the corresponding two base pairs per ``rung," as well
as the poorly-conducting phosphate backbone chain, can all be
reduced into an effective 1D model with renormalized onsite energy
and hopping constant.\c{back} The effective constants are those
obtained from first-principles calculations which estimate
bandwidths in these systems.\c{endres}}

When the concentration of one type of site is small, say $B$, the
probability for two nearby sites to have the same onsite energy
$\va_B$ is smaller. In this case, the system tends to the
``repulsive binary alloy" model, in which one extended state
exists.\c{ram} A simple calculation yields the transmission
coefficient for a system  with one impurity with onsite energy
$\va_B$,
 \be
 T_1(E)=\f{(2t_{AB}^2 \sin k)^2}{(2t_{AB}^2 \sin k)^2+N_1^2}\, ,
 \label{T_1}
 \ee
where $N_1=Wt_{AA}+2(t_{AB}^2-t_{AA}^2)\cos k$, $E=2t_{AA}\cos k$,
and $W=\va_B-\va_A$. One can see that for the state with energy
$E=Wt_{AA}^2/(t_{AA}^2-t_{AB}^2)$, the transmission coefficient is
unity. The states {\em near} this energy have large transmission
coefficient and long localization length, even in systems with
more $B$ impurities. In fact, these states have an important
contribution to transport. Figure 1 shows the transmission
coefficient for various concentration of $B$ impurities with
energy $\va_B$. The transmission is obtained by a transfer matrix
calculation for 1000 sites, and averaged over 300 different
configurations. For the purpose of comparison, we also show the
transmission coefficient for the Anderson model (where
$t_{AA}=t_{BB}=t_{AB}=t$) with the same degree of onsite disorder.
We see that in {\em all} cases, the local correlation built-in
through the $t$ values leads to {\em much larger} transmission
coefficient, compared with those in the Anderson model. When the
concentration is small, there is in fact a regime of high
transmission ($\sim$ 1), with energy centered around that given by
Eq.\ (\ref{T_1}).

\begin{figure}[b]
\includegraphics[width=8cm]{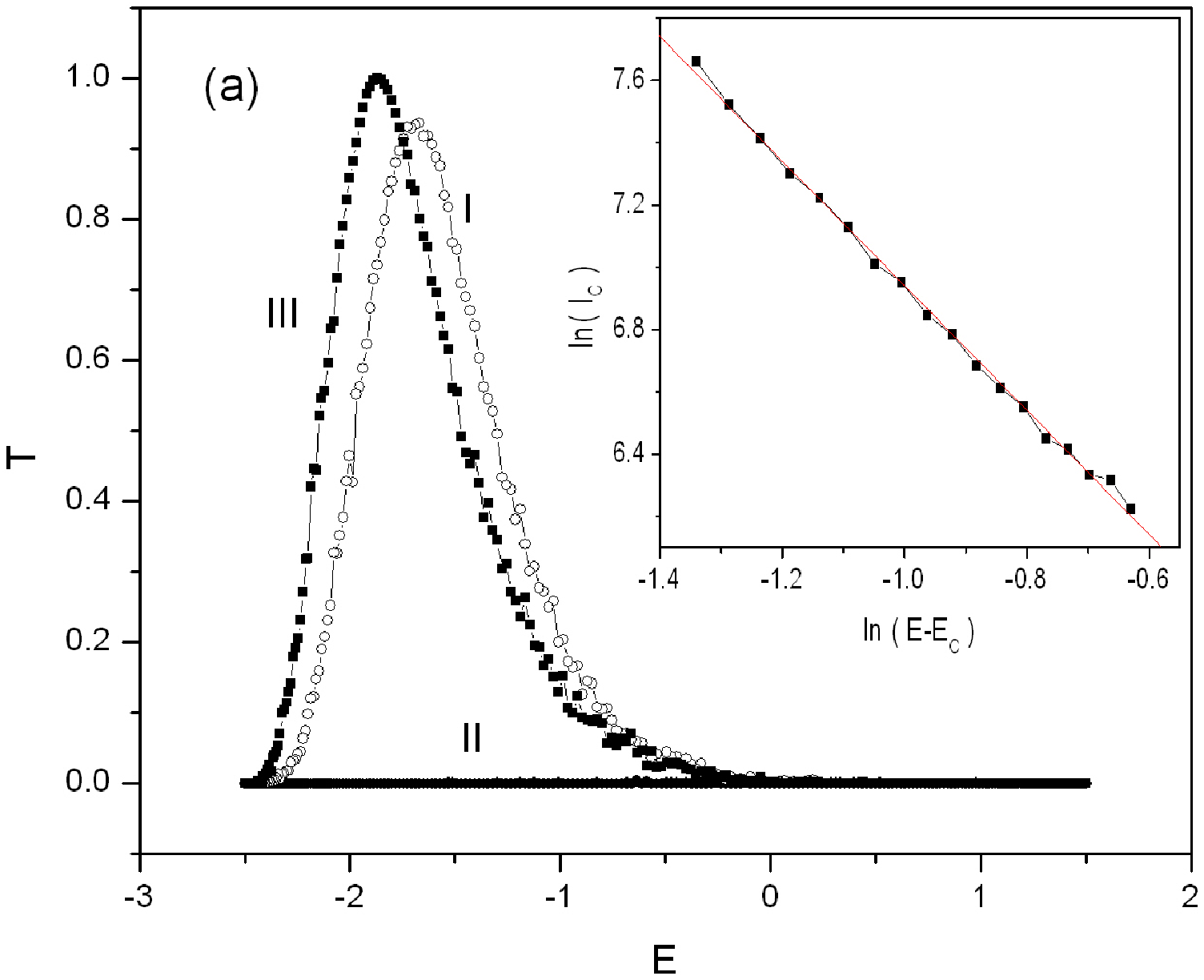}
\includegraphics[width=8cm]{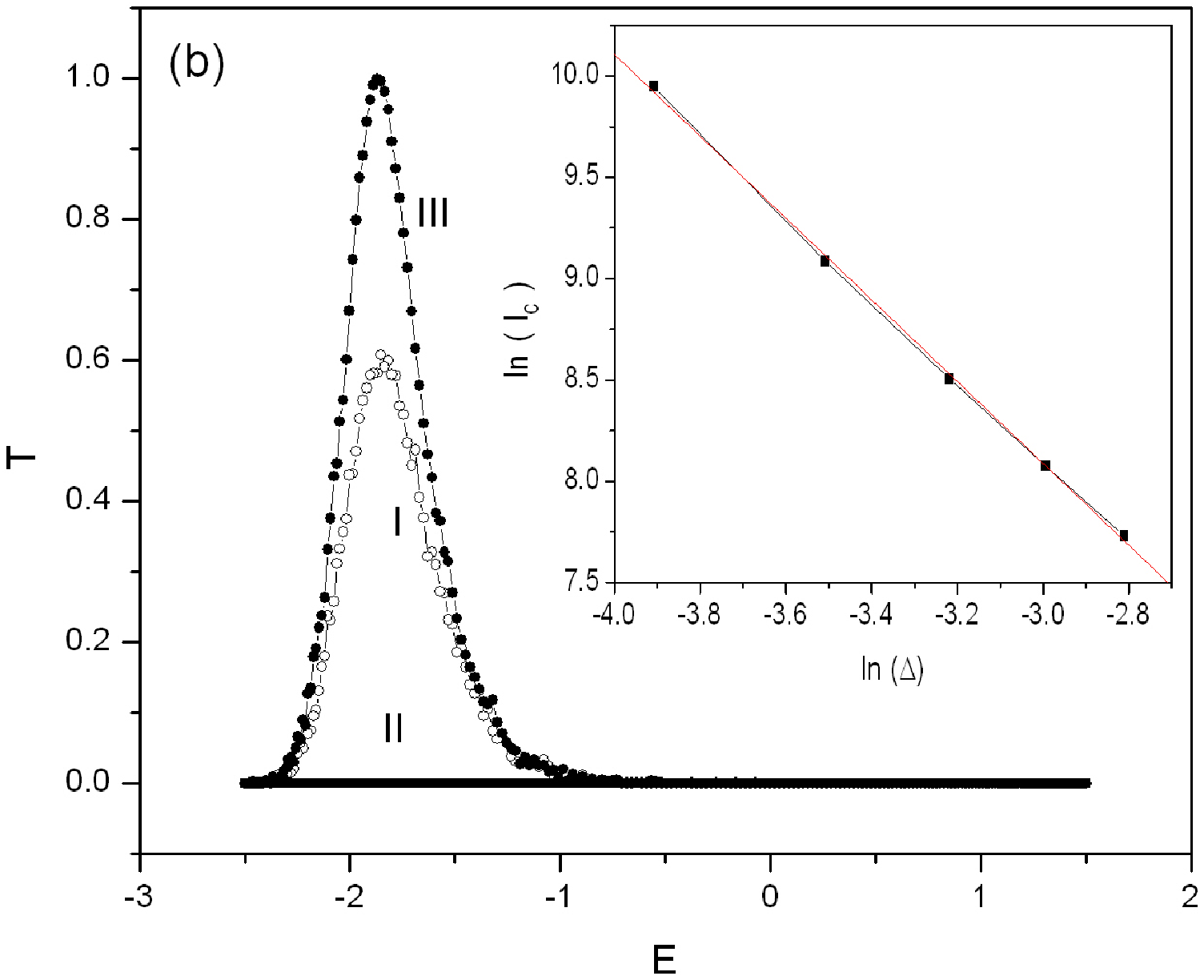}
\caption{(a) Transmission coefficients vs.\ energy;
$W=\va_B-\va_A=1$. Curve I (empty symbols) is typical for system
with local correlation; hopping constants here $t_{AA}=1$,
$t_{BB}=1.73$, $t_{AB}=(t_{AA}+t_{BB})/2=1.36$. Curve II (solid
line near zero) is for Anderson limit, with all hopping constants
equal ($=1$). Curve III (solid symbols) is for system with
``golden correlation", $t_{AB}= t_G \equiv \sqrt{t_{AA}t_{BB}}
=1.316$, with unit transmission at $E\simeq -1.9$. Concentration
of $\va_{B}$ site is 0.1 in all three curves.
 (b) Same as in (a), but with concentration of $\va_{B}$ at 0.5.
 Inset in (a): Localization length $l(E)$ vs.\ $(E-E_c)$ for system with golden
correlation. Slope of fitted line is 2. Concentration of $\va_B$
site is 0.5.
 Inset in (b): Localization length at critical energy $l_c=l(E_c)$ vs.\
$D=(|t_{AB}-t_G|)$, where $t_G=\sqrt{t_{AA}t_{BB}}$ is the golden
condition. Concentration of $\va$ is 0.5. }
\end{figure}

With increasing $B$ concentration, the electron has a higher
chance to scatter from dimer and trimer impurities.  For a single
dimer impurity, a straightforward but cumbersome calculation
yields the transmission
 \be
T_2(E)=\f{(2t_{AA}t_{BB}t_{AB}^2 \sin k)^2}{(2t_{AA}t_{BB}t_{AB}^2
\sin k)^2+N_2^2} \, ,
 \ee
where $N_2=(t_{AA}^2-t_{AB}^2)(W-2t_{AA}\cos k)^2 +%
t_{AB}^4-t_{AA}^2 t_{BB}^2 + t_{AB}^2 W^2-2Wt_{AA} t_{AB}^2 \cos k%
$. In general, there are more energy values satisfying $T_2(E)=1$,
but they are different than $T_1(E)=1$, in general. Consequently,
although short-range off-diagonal correlation leads to extended
states for finite systems (localization length $l_c$ larger than
the system size $L$), these states are not extended states in the
thermodynamic limit, $L\rightarrow \infty$.

Figure 1 also illustrates that for special local correlations
there is an extended state even for infinite systems. One can
easily verify that when $t_{AB}=t_G \equiv \sqrt{t_{AA}t_{BB}}$
(we call this the ``golden correlation" in off-diagonal
parameters), the condition $T_1(E)=T_2(E)=1$ can be satisfied.
This implies that for the peculiar golden correlation $t_G$,
single and dimer impurities are essentially transparent at this
energy. The question remains of how general is this result, that
is, how about trimers or more general impurities? Instead of
calculating $T_m(E)$ for $m$ impurities, we prove the existence of
an extended state by explicit construction. It is not difficult to
check that under the condition $t_{AB}=t_G$, the state $\al_n
e^{ikn}$, with $\al_n=1$, for $\va_n=\va_A$, and
$\al_n=\sqrt{{t_{AA}}/{t_{BB}}}$, for $\va_n=\va_B$, is indeed an
extended state with energy $E=E_c=\va_A+2t_{AA}\cos
k=\va_B+2t_{BB}\cos k$. The physical picture for this state is
then that the electron propagates on island $A$ or $B$ in the
plane wave form, while the golden condition ensures perfect
transition from island $A$ to island $B$, and vice versa.  One can
say that this perfect transmission arises from the cancellation of
backscattered waves produced by the subtle tuning of off-diagonal
correlations. We find a state with unit transmission coefficient
as that shown by curve III in Fig.\ 1(a), {\em even for high
concentration of impurities}, although this ``resonance" becomes
sharper for high impurity concentrations (see Fig.\ 1(b)). This is
the first example of an extended state in the thermodynamic limit
in a random 1D system with short-range off-diagonal correlations
(but {\em no} correlation in onsite energies). Notice also that
$T(E_c)=1$ even for a system with 50\% disorder, as shown in Fig.\
1(b).

Under the golden correlation condition $t_{AB}=t_G$, the extended
state satisfies $2\cos k=(\va_B-\va_A)/(t_{AA}-t_{BB})$, which can
be met only when $|\va_B-\va_A|<2|t_{AA}-t_{BB}|$, resulting in an
interesting effect. Usually in the presence of only diagonal or
off-diagonal disorder, the larger the disorder is, the poorer is
the transport. The situation is quite different for correlated
diagonal and off-diagonal disorder. To obtain an extended state in
the presence of the diagonal difference $W=\va_B-\va_A$, the
difference between $t_{AA}$ and $t_{BB}$ has to be large enough,
i.e., one needs the correlated off-diagonal disorder to be {\em
large}. This is contrary to expectations.

Notice also that for fixed $t_{AA}$, and $t_{BB}$, there is a
critical onsite difference $W=2|t_{AA}-t_{BB}|=2\De$. From the
time evolution of a particle initially placed at a randomly chosen
site (not show here), we find that when $W<2\De$, the mean square
displacement in time $\tau$ is $\langle x^2 \rangle \sim
\tau^{3/2}$, and then it is in a superdiffusive phase. In
contrast, when $W=2\De$, the system is in a diffusive phase,
$\langle x^2 \rangle \sim \tau$; and for $W>2\De$, the mean square
displacement is bounded. This transition is similar to that in the
random dimer model (RDM), although with different
characteristics.\cite{ram} In RDM, the transition occurs at
$W=2t_{AA}$ (all $t$ the same). In our case, the condition is
related to the difference between hopping constants, and not the
hopping constants themselves. It is interesting that for
$t_{AA}=t_{BB}<W/2$, there are extended states in RDM, but no
extended state in our model.

We study the localization length $l(E)$ for states near the
critical energy $E_c$ in the Fig.\ 1(a) inset.  We find that $l(E)
\propto (E-E_c)^{-2}$ for states near $E_c$. The number of
extended states for a system of length $L$ (i.e., $l(E)>L$) is
related to $\de k \propto E-E_c \propto L^{-1/2}$, where near
$E_c$, $E=E_c+A\de k$. The number of extended states is then $\de
k/ (1/L)=L^{1/2}$, a sizable number, just as in the RDM.
\footnote{Note that near the band edge, there is an anomaly, and
$(\de k)^2 \propto E-E_c \propto L^{-1}$, but the number of
extended states is still $\propto L^{1/2}$. }

The long time behavior of the system is determined by a critical
exponent. One can show the relation between two exponents $\theta$
and $\g$, defined by $\langle x^2 \rangle \sim \tau^{\theta}$, and
$l(E)\sim |E-E_c|^{-\g}$. For short times, the electron has
ballistic behavior, since it has not sampled yet the disorder
potential, so that $\langle x^2 \rangle \sim (v\tau)^2$. For long
time, however, $\langle x^2 \rangle \sim l^2(E)$, for an electron
with energy $E$. We can then write the mean square displacement as
$\langle x^2 \rangle=\int dE \, \rho(E) (v\tau)^2 \, f \left(\f{v%
\tau}{l(E)} \right) \, $, where $\rho (E)$ is the density of
states, and we surmise the scaling function $f(x) \rightarrow 1$,
as $x\rightarrow 0$, and $f(x) \rightarrow 1/x^2$, as
$x\rightarrow \infty$.  From this, one obtains $\langle x^2
\rangle\sim \tau ^{2-1/ \g}$, for long times, so that
$\theta=2-1/\g$. When $\g=2$, as in Fig.\ 1(a) inset,
$\theta=\f{3}{2}$ (superdiffusive regime); while when $\g=1,
\theta=1$ (diffusive). There is perfect agreement with our
numerical calculations.

It is natural to expect that in many systems there exist
correlations between diagonal and off-diagonal disordered
parameters. However the golden correlation condition is not
necessarily satisfied, and it is important to see how the
transport properties change when a system deviates from this. The
inset in 1(b) shows that $l(E_c)\propto (t_{AB}-t_G)^{-2}$, so
that to obtain extended states, we need $|t_{AB}-t_G|<L^{-1/2}$.
As long as this condition is met, effective conduction channels
are opened by the off-diagonal correlations in the disordered
system.

Our predictions could be verified experimentally in systems with
access to varying degree of disorder and structural correlation,
such as model semiconductor superlattices ({\bf SL}s). \c{lat}
Consider a SL with quantum wells of two different widths $d_A$ and
$d_B$, distributed randomly in the structure. The barriers between
wells have the same height $U$ (given by the material composition)
and width $b_A$ (or $b_B$) if the barrier is between two alike
wells of width $d_A$ (or $d_B$), and otherwise have width $b_C$.
An estimate of the hopping constant between two quantum wells with
width $d_L$ and $d_R$, separated by a barrier of width $b$ and
height $U$ is $t=\f{\pi^2\hbar^2}{ms}\sqrt{d_L d_R} \exp ({-sb})$,
where $s=\sqrt{2mU}/\hbar$. By tuning parameters, the golden
condition can be attained.  Figure 2 shows the transmission for
different systems calculated from a Kronig-Penney model of the
SL.\@ Curve A is for a system satisfying the golden condition, as
estimated from the expression above, while curves B and C are
results away from the condition. The discussion above for the
tight-binding model suggests that transport would indeed be better
for system in curve A, even as the barrier between different
quantum wells (curve B) is {\em thinner}. We emphasize that Fig.\
2 is obtained from a Kronig-Penney model of the structure, so that
hopping amplitudes go far beyond nearest neighbors, and the golden
condition is likely much more involved than in the tight-binding
model.  The golden condition for curve A was {\em not} optimized,
but just estimated from the relation above, and the difference
between these curves is remarkable. \footnote{Despite averaging
over 600 configurations, curve A in Fig.\ 2 exhibits large
oscillations for $E\simeq E_c$, reflecting the sensitivity of
these states to boundaries and disorder, which produces poor
self-averaging and arises from their extended nature.}

\begin{figure}[b] \vspace{-1em}
\includegraphics[angle=90,width=9cm]{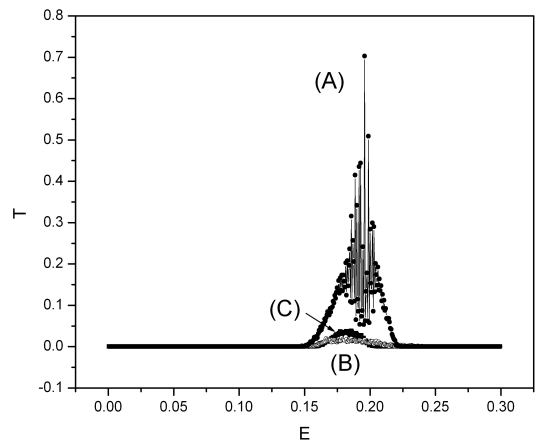}
\caption{Transmission for SL with 100 randomly distributed quantum
wells of two types, width 2.6 (type {\sf a} well) and 2.9nm (type
{\sf b}). Barriers between same {\sf a} (or {\sf b}) wells have
3.6nm (2.4nm) width. Other barrier width is 3.0nm (curve A, $\sim$
golden condition), 2.0nm (curve B), and 3.8nm (C). All barriers
have height 0.3 eV.\@ $T(E)$ averaged over 600 different disorder
configurations.  Concentration of {\sf b} wells is 0.5.}
\end{figure}

As discussed before, our studies have direct application to models
of transport in DNA in the literature.\c{back,models} For a
typical DNA molecule the base pair sequence may be essentially
random, such as in $\la$-DNA. However, the chemical structure
determines the local correlation between onsite energies and
hopping constant via the $\pi-$orbital overlap. In order to
explore how the local correlation changes transport, we compare
the I-V curves of different systems,  obtained using the
Landauer-B\"uttiker formalism, \c{landauer} $I=(2e/h)\int dE\,
T(E)\, [f_L(E)-f_R(E)]$, where $f_{L/R}(E)
=\{\exp[E-\mu_{L/R}/k_BT]+1\}^{-1}$ is the Fermi function. We
choose $\mu_L=E_F+(1-\kappa )eV$, and $\mu_R=E_F-\kappa eV$, where
$E_F$ is the equilibrium Fermi energy, $V$ is the applied voltage,
and $\kappa$ is a parameter describing the possible asymmetry of
contact to leads, chosen here as $\kappa=1/3$. \c{deph,Datta} We
assume that the DNA is attached to ideal leads described by a
metal with bandwidth 1.2eV.\@ The hopping constant between leads
and DNA chain is chosen to be $\sim t_{AB}/10 \simeq 0.01$eV,
reflecting a relatively poor contact. We use two different sets of
parameters: $t_{AA}=-0.0695$eV, $t_{BB}=-0.1409$eV, and
$t_{AB}=(t_{AA}+t_{BB})/2$, curve A in Fig.\ 3, describe a
realistic molecule, as the values are obtained from microscopic
calculations; \c{endres} $t_{AA}=t_{BB}=t_{AB}=-0.1403$eV, curve
B, simulates an uncorrelated system, i.e., the Anderson limit.

We can see from Fig.\ 3 that the current in the system with local
correlation (curve A) is overall much larger than in the system
without correlation, even though the hopping constant is larger in
B.\@ There is in fact no conductance over the entire bias range
for curve B, with no correlation in the hopping constants. The
message of these results is that even for DNA with random
sequences, such as  $\la$-DNA, ``good" transport is possible due
to the effective conduction channels opened by structural
correlations. Notice that $t_{AB}$ in curve A does not satisfy the
golden condition ($\sim -0.099$eV) by about 4\%, and yet, there is
significant current amplitude for finite biases. In contrast to
the conducting states in polymers, which arise from the
correlation in local energies, the conducting states here have to
do with correlation in hopping amplitudes. It is clear that the
backbone may change the local correlations. We may conclude that
changes in local correlation will lead to changes in the I-V
features, which may in fact be an ingredient in recent
experiments, especially if chemical changes affect the molecule
structure. \c{hartzell}

\begin{figure}[b]  
\includegraphics[angle=90,width=10cm]{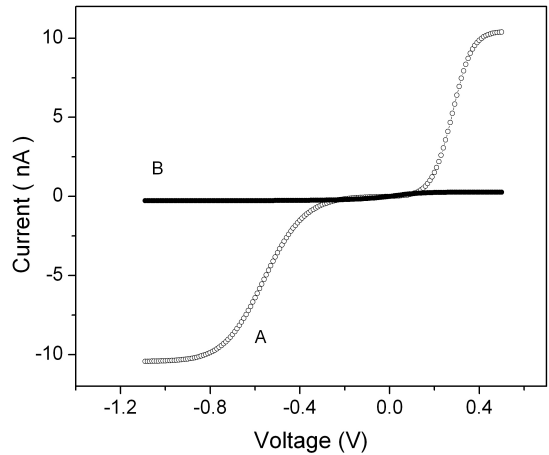}
\caption{I-V curves for a random base pair sequence (i.e., random
onsite energies). Curve A is for model of $\la$-DNA with realistic
local correlation in the hopping amplitudes. Curve B is for random
diagonal Anderson model with hopping amplitudes set equal. Size of
systems is 562; temperature in Fermi broadening is 300K.}
\end{figure}

We acknowledge support from DOE grant no.\ DE-FG02-91ER45334, NSF
NIRT grant no.\ 0103034, and the CMSS Program, and discussions
with the NIRT group at Ohio U.

\end{document}